\documentclass{iopart}
\usepackage{amssymb,graphicx}

\newcommand{\dd}{\mathrm{d}}
\newcommand{\mean}[1]{\left\langle #1 \right\rangle}

\newcommand{\pd}[2]{\frac{\partial #1}{\partial #2}}

\newcommand{\IInt}[3]{\int_{#2}^{#3}\dd #1\;}
\newcommand{\Int}[1]{\int\dd #1\;}

\newcommand{\ctrl}{\alpha}
\newcommand{\lam}{\lambda}

\newcommand{\ps}{p_\mathrm{s}}
\newcommand{\js}{j_\mathrm{s}}
\newcommand{\vs}{\nu_\mathrm{s}}
\newcommand{\tot}{_\mathrm{tot}}
\newcommand{\dis}{_\mathrm{dis}}
\newcommand{\ex}{_\mathrm{ex}}
\newcommand{\hk}{_\mathrm{hk}}
\newcommand{\eps}{\varepsilon}
\newcommand{\pr}{\mathrm{prob}}

\newcommand{\FP}{\mathcal L}
\newcommand{\OA}{\mathcal A}

\newcommand{\OP}{\mathcal P}

\begin{document}

\letter{Integral fluctuation theorem for the housekeeping heat}
\author{T. Speck and U. Seifert}
\address{{II.} Institut f\"ur Theoretische Physik, Universit\"at Stuttgart,
  70550 Stuttgart, Germany}
\ead{speck@theo2.physik.uni-stuttgart.de}

\begin{abstract}
  The housekeeping heat $Q\hk$ is the dissipated heat necessary to maintain
  the violation of detailed balance in nonequilibrium steady states. By
  analyzing the evolution of its probability distribution, we prove an integral
  fluctuation theorem $\mean{\exp[-\beta Q\hk]}=1$ valid for arbitrary driven
  transitions between steady states. We discuss Gaussian limiting cases and
  the difference between the new theorem and both the Hatano-Sasa and the
  Jarzynski relation.
\end{abstract}

{\sl Introduction. --} Steady state thermodynamics provides a framework for
describing transitions between different nonequilibrium steady states using
terms familiar from ordinary thermodynamics~\cite{oono98,hata01,sasa04}. As a
crucial concept, the division of the total heat dissipated in such transitions
into two contributions, a housekeeping heat $Q\hk$ and an excess heat $Q\ex$,
has emerged. The housekeeping heat is the one permanently dissipated while
maintaining a nonequilibrium steady state at fixed external parameters
$\ctrl$. The excess heat is the one associated with a transition between
different steady states caused by changing $\ctrl$. For transitions between
equilibrium states the housekeeping heat vanishes and the total heat reduces
to the excess heat. These concepts have been made explicit in the model of a
colloidal particle driven along a one-dimensional coordinate described by a
Langevin equation ~\cite{hata01,seki98}. In particular, it was shown that the
excess heat $Q\ex$ obeys a fluctuation theorem
\begin{equation}
  \mean{\exp[-\beta Q\ex-\Delta\phi]} = 1,
  \label{eq:hs}
\end{equation}
where $\phi(x,\ctrl)\equiv-\ln\ps(x,\ctrl)$ with $\ps$ the steady state
probability distribution for fixed external parameters $\ctrl$. The brackets
$\mean{\cdots}$ denote an average over many different realizations of the
process in a surrounding heat bath of inverse temperature
$\beta\equiv1/k_\mathrm{B}T$. Since the average $\mean{\Delta\phi}$ can be
identified as the change in generalized system entropy $\Delta S$, the
relation $\Delta S\geqslant-\beta\mean{Q\ex}$ following from
equation~(\ref{eq:hs}) corresponds to a generalization of the second law to
transitions between two different steady states~\cite{hata01}.  An
experimental test of this fluctuation theorem was realized using laser
tweezers dragging a colloidal particle through a viscous fluid~\cite{trep04}.
In this Letter, we prove a similar fluctuation theorem for the housekeeping
heat
\begin{equation}
  \mean{\exp[-\beta Q\hk]} = 1
  \label{eq:ft}
\end{equation}
valid for arbitrarily driven systems described by such a Langevin equation.

Integral fluctuation theorems of the type $\mean{\exp[-A]}=1$ have been
derived and the corresponding probability distribution $P(A=a)$ has been
discussed for a variety of systems showing an underlying stochastic
dynamics~\cite{jarz97,jarz97a,croo00,maes03,rito04,seif04,spec04,seif05a,impa05}.
The most prominent example is the Jarzynski relation~\cite{jarz97}, where $A$
is the dissipated work spent driving the system from one equilibrium state to
another. Crucial for the Jarzynski relation proper is the fact that the system
obeys detailed balance for fixed external parameters $\ctrl$. However, even if
this is not the case, by applying the same reasoning a Jarzynski-like relation
\begin{equation}
  \mean{\exp[-\beta Q\tot-\Delta\phi]} = 1
  \label{eq:jar}
\end{equation}
can be derived, which differs from equation~(\ref{eq:hs}) by using the total
heat $Q\tot$ as shown below. It is crucial to appreciate that all three
relations~(\ref{eq:hs}-\ref{eq:jar}) are genuinely different and require a
different derivation.

Such integral fluctuation theorems should be distinguished from the detailed
fluctuation theorem $P(-a)/P(+a)=e^{-a}$ valid for a probability distribution
$P(a)$ in steady states at constant external parameters. Such a theorem was
discussed first for entropy production in sheared two-dimensional
fluids~\cite{evan94,gall95} and later generalized to chaotic and stochastic
dynamics~\cite{kurc98,lebo99,gasp04,seif05}. It even holds for periodically
driven systems as demonstrated experimentally using a single two-level
system~\cite{schu05}.

\begin{figure}[t]
  \centering
  \includegraphics[width=6cm]{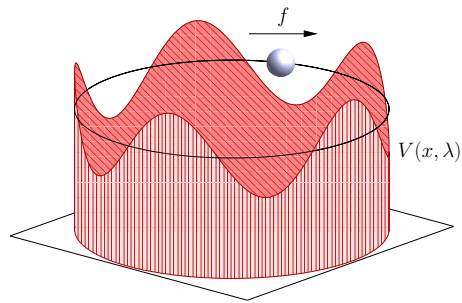}
  \caption{A periodic potential $V(x,\lam)$ with period $L=2\pi$, where $x$ is
    the angular position of a driven colloidal particle moving through a
    viscous fluid representing the surrounding heat bath. Depth and shape of
    the potential depend on the externally controlled parameter $\lam$. A
    force $f$ can be applied directly to the particle.}
  \label{fig:ring}
\end{figure}

{\sl The model. --} As paradigm, we will focus on driven diffusive motion of a
single colloidal particle in one dimension. The potential
$V(x,\lam)=V(x+L,\lam)$ is spatially periodic with period $L$ and depends on
an external parameter $\lam$, see figure~\ref{fig:ring}. The Langevin equation
for the position $0\leqslant x(t)<L$ of the particle becomes
\begin{equation}
  \gamma\dot x = -V' + f + \eta,
\end{equation}
where $V'\equiv\partial V/\partial x$ and $f(t)$ is a time-dependent
nonconservative force. The thermal noise $\eta$ obeys the usual correlations
$\mean{\eta(t)\eta(t')}=2(\gamma/\beta)\delta(t-t')$ with friction coefficient
$\gamma$. The externally controllable parameters of the system are
$\alpha\equiv(\lam,f)$. A transition from $\ctrl_0\equiv\ctrl(0)$ to
$\ctrl_t\equiv\ctrl(t)$ occurs along the path $\ctrl(\tau)$ through parameter
space.

Following Sekimoto~\cite{seki98}, we write the total exchanged heat between
system and reservoir in the time interval $0\leqslant\tau\leqslant t$ as the
functional
\begin{equation}
  Q\tot[x(\tau);\ctrl(\tau)] \equiv \IInt{\tau}{0}{t} \dot x(\tau)
  \left[f(\tau)-V'(x(\tau),\lam(\tau))\right].
  \label{eq:heat}
\end{equation}
We take the sign of the heat to be positive when energy flows out of the
system into the heat bath.

In the case of genuine nonequilibrium states where $\ps(x,\ctrl)$ violates
detailed balance at fixed $\ctrl$ (in our model for $f\neq0$) it makes sense
to split up the total heat into the housekeeping heat, i.e., the part needed
to maintain the violation of detailed balance
\begin{equation}
  Q\hk[x(\tau);\ctrl(\tau)] \equiv 
  \gamma\IInt{\tau}{0}{t} \dot x(\tau)\vs(x(\tau),\ctrl(\tau)),
  \label{eq:Qhk}
\end{equation}
and the excess heat $Q\ex\equiv Q\tot-Q\hk$~\cite{hata01}. The function $\vs$
is the local mean velocity of the particle in the steady state at fixed
$\ctrl$,
\begin{equation}
  \vs(x,\alpha) \equiv \frac{\js(\alpha)}{\ps(x,\alpha)},
\end{equation}
where $\js$ is the steady state probability current.

{\sl Evolution equation. --} In the definition~(\ref{eq:Qhk}) of the
housekeeping heat the singular velocity $\dot x$ appears, which requires a
regularization. We therefore introduce the momentum $p=m\dot x$ of the
particle with mass $m$ as an additional degree of freedom, even though we are
considering the overdamped case. Then the coupled Langevin equations become
\begin{eqnarray}
  \dot x &=& p/m, \\
  \dot p &=& f - V' - (\tilde\gamma/m)p + \zeta \quad\textrm{with}\quad
  \mean{\zeta(t)\zeta(t')}=2(\tilde\gamma/\beta)\delta(t-t'), \\
  \label{eq:heat:v}
  \dot q &=& (\gamma/m)p\vs.
\end{eqnarray}
In equation~(\ref{eq:heat:v}), $\dot q$ is the dissipation rate of the
housekeeping heat (\ref{eq:Qhk}) with $\dot x$ replaced by $p$. The joint
probability $\bar\rho(x,p,q,t)$ to find the particle at position $x$ with
momentum $p$ and to have spent a housekeeping heat $q$ up to time $t$ obeys
the generalized Fokker-Planck equation
\begin{equation}
  \partial_t\bar\rho = \left[ -\frac{p}{m}\partial_x + (V'-f)\partial_p +
  \eps\partial_p p + \eps\frac{m}{\beta}\partial_p^2 - 
  \frac{\gamma}{m}p\vs\partial_q \right] \bar\rho.
  \label{eq:barrho}
\end{equation}
Since $Q\hk$ was defined for overdamped motion, we fix the friction
coefficient $\gamma$ and use $\eps\equiv\tilde\gamma/m$ as expansion parameter
to recover the appropriate dynamics for $Q\hk$ in the limit of large $\eps$.

Adiabatic elimination of fast variables is a standard technique in stochastic
dynamics~\cite{risken}. To simplify notation we introduce the two differential
operators
\begin{eqnarray}
  \FP_p &\equiv& \partial_p[p+(m/\beta)\partial_p], \\
  \label{eq:A}
  \OA   &\equiv& (p/m)\partial_x - (V'-f)\partial_p + (\gamma/m)p\vs\partial_q,
\end{eqnarray}
where we use the usual convention that a differential operator like
$\partial_x$ acts on all factors placed to its right. Using these operators, we
rewrite equation~(\ref{eq:barrho}) as
\begin{equation}
  \partial_t\bar\rho = \left[\eps\FP_p-\OA\right]\bar\rho.
\end{equation}
Further we define a projector by 
\begin{equation}
  \OP\psi(x,p) \equiv \psi_0(p)\Int{v}\psi(x,p),
\end{equation}
where the Maxwell distribution $\psi_0(p)=\sqrt{\beta/2\pi m}\,e^{-\beta
  p^2/2m}$ is the stationary solution of $\FP_p\psi_0=0$. This ensures the two
important properties
\begin{equation}
  \OP\FP_p = \FP_p\OP = 0,\quad \OP\OA\psi_0 = 0
  \label{eq:pro}
\end{equation}
of the projector $\OP$. Now we decompose the joint probability
$\bar\rho=\bar\rho_0+\bar\rho_1$ into a part ``parallel'' to the stationary
distribution
\begin{equation}
  \bar\rho_0 \equiv \OP\bar\rho = \rho\psi_0
  \label{eq:barrho:0}
\end{equation}
and a part perpendicular, $\bar\rho_1\equiv(1-\OP)\bar\rho$. The function
\begin{equation}
  \rho(x,q,t) = \Int{p} \bar\rho(x,p,q,t)
\end{equation}
is the probability distribution we are looking for. Physically, this scheme
corresponds to a decomposition into a part for which position $x$ and momentum
$p$ are uncorrelated and into a deviation $\bar\rho_1$. For uncorrelated
stochastic variables the joint probability decays to the simple product
(\ref{eq:barrho:0}).

Taking advantage of the properties (\ref{eq:pro}) the coupled equations of
motion become
\begin{eqnarray}
  \label{eq:para}
  \partial_t\bar\rho_0 &=& -\OP\OA\bar\rho_1, \\
  \label{eq:perp}
  \partial_t\bar\rho_1 &=& \eps\FP_p\bar\rho_1 - \OA\bar\rho_0 -
  (1-\OP)\OA\bar\rho_1.
\end{eqnarray}
The second equation~(\ref{eq:perp}) can be expanded into powers of
$\eps^{-1}$, yielding up to the first order $\bar\rho_1 \simeq
\eps^{-1}\FP_p^{-1}\OA\bar\rho_0$. Putting this back into
equation~(\ref{eq:para}) we finally arrive at
\begin{equation}
  \partial_t\rho = -\eps^{-1}\left[ \Int{p}\OA\FP_p^{-1}\OA\psi_0 \right] \rho.
\end{equation}
For the calculation we insert definition~(\ref{eq:A}) and note that
$\FP_p^{-1}p\psi_0=-p\psi_0$. Gaussian integration then leads
straightforwardly to the evolution equation
\begin{equation}
  \partial_t\rho = \left[ \FP +
    \left(2\partial_x\vs-\frac{\vs^2}{D}\right)\frac{\partial_q}{\beta} +
    \frac{\vs^2}{D}\frac{\partial_q^2}{\beta^2} \right] \rho,
  \label{eq:rho:q}
\end{equation}
where we have set $\tilde\gamma=\gamma$ and used the Einstein relation
$D=1/\gamma\beta$. The operator
\begin{equation}
  \FP \equiv \partial_x\left[(V'-f)/\gamma+D\partial_x\right]
\end{equation}
is the usual Fokker-Planck operator. The probability distribution of the
housekeeping heat
\begin{equation}
  P(q,t) = \Int{x}\rho(x,q,t)
\end{equation}
can then be calculated simply by integrating over $x$. Since we start in a
steady state, the initial condition needed to solve equation~(\ref{eq:rho:q})
is $\rho(x,q,0)=\ps(x,\ctrl_0)\delta(q)$.

{\sl Proof of the fluctuation theorem~(\ref{eq:ft}). --} In the time
derivative
\begin{equation}
  \partial_t\mean{\exp[-\beta Q\hk]} = \Int{x\dd q} e^{-\beta q}
  \partial_t\rho(x,q,t)
\end{equation}
we insert the equation of motion (\ref{eq:rho:q}) and get rid off the partial
derivatives with respect to $q$ by integration by parts with vanishing
boundary terms. We then have
\begin{equation}
  \partial_t\mean{\exp[-\beta Q\hk]} = \Int{x\dd q} e^{-\beta q}
  \rho(x,q,t)[-\vs^2+\vs^2]/D = 0.
\end{equation}
Therefore the average over trajectories of length $t$ becomes
\begin{equation}
  \mean{\exp[-\beta Q\hk]}_t = \mean{\exp[-\beta Q\hk]}_0 = 1,
\end{equation}
since for $t=0$ we start with $Q\hk=0$. This proof of the fluctuation
theorem~(\ref{eq:ft}) for any time-dependence $\ctrl(\tau)$ is the central
result of this Letter.

{\sl Mean and variance. --} In order to obtain an explicit expression for the
mean of $Q\hk$, we integrate equation (\ref{eq:rho:q}) with $\Int{x\dd q}q$
and obtain after integration by parts with respect to $q$
\begin{equation}
  \partial_t\mean{Q\hk} = \gamma\Int{x}\vs^2(x,\ctrl)p(x,t) 
  = \gamma\mean{\vs^2(\ctrl)} \geqslant 0
\end{equation}
and hence
\begin{equation}
  \mean{Q\hk} = \gamma\IInt{\tau}{0}{t}\mean{\vs^2(\ctrl(\tau))} \geqslant 0.
  \label{eq:mean}
\end{equation}
In general, the average over $\vs^2$ has to be taken using the time-dependent
probability $p(x,t)$ of the position $x$. However, in a steady state we have
$p(x,t)=\ps(x,\ctrl)$ and the mean becomes
\begin{equation}
  \mean{Q\hk} = \gamma\mean{\vs^2}t = \gamma\left[\js^2\Int{x}\ps^{-1}\right]t,
  \label{eq:mean:ss}
\end{equation}
leading to a constant mean dissipation rate.

Applying the same scheme to the second moment leads to a variance
\begin{equation}
  \sigma^2 = 2\gamma\IInt{\tau}{0}{t}\mean{\vs^2(\ctrl(\tau))Q\hk(\tau)} 
  + (2/\beta)\mean{Q\hk} - \mean{Q\hk}^2.
  \label{eq:var}
\end{equation}
The variance thus involves weighting the housekeeping heat with the square
velocity $\vs^2$ in the first term. This recursive scheme can be continued to
obtain any moment or cumulant as a function of all previous moments or
cumulants and the weight function $\vs^2$.

{\sl Limiting cases. --} In the limiting case $V'\ll f$ which corresponds to
the situation where $f$ is large such that the particle hardly ``feels'' the
underlying potential we can solve equation~(\ref{eq:rho:q}) exactly. Then the
stationary distribution becomes uniform and $\vs(\tau)\approx f(\tau)/\gamma$.
By integrating over $x$ in equation~(\ref{eq:rho:q}) we get for the probability
distribution $P(q,t)$ the evolution equation
\begin{equation}
  \partial_t P = -\gamma^{-1}f^2\partial_q P + Df^2\partial^2_q P.
\end{equation}
This is a diffusion-like equation and its solution is a Gaussian with mean
\begin{equation}
  \mean{Q\hk} = \gamma^{-1}\IInt{\tau}{0}{t}f^2(\tau)
\end{equation}
and variance $\sigma^2=(2/\beta)\mean{Q\hk}$, which obviously obeys the
integral constraint implied by the fluctuation theorem~(\ref{eq:ft}).

In the opposite case $V'\gg f$ for $\dot\ctrl=0$ the local mean velocity $\vs$
is small, which we make explicit by $\vs\rightarrow\eps\bar\vs$. Now a time
scale separation using $\eps$ as a small parameter becomes possible. The fast
time variable is simply $t$ whereas the slow time variable $s\equiv\eps t$ is
given by the ``hopping time'', i.e., the mean time the particle needs to
complete one period. Switching to the slow time scale $\rho(x,q,s)$,
equation~(\ref{eq:rho:q}) becomes
\begin{equation}
  \partial_s\rho = \left[ \eps^{-1}\FP +
    \left(2\partial_x\bar\vs-\eps\frac{\bar\vs^2}{D}\right)
    \frac{\partial_q}{\beta} +
    \eps\frac{\bar\vs^2}{D}\frac{\partial_q^2}{\beta^2} \right] \rho.
\end{equation}
In order to eliminate $x$ for small $\eps$, we apply exactly the same scheme
developed above for elimination of the momentum $p$ for large friction
$\gamma$. Again dividing the distribution $\rho$ into a part parallel to the
stationary distribution $\rho_0(x,q,s)=P(q,s)\ps(x)$ and into a part $\rho_1$
perpendicular, we see that $\rho_1$ is of order $\eps^2$. If we take into
account only terms of order $\eps$ and restore the original variables, we get
the evolution equation
\begin{equation}
  \partial_t P = -\gamma\mean{\vs^2}\partial_q P 
  + (\gamma/\beta)\mean{\vs^2}\partial^2_q P.
  \label{eq:lin}
\end{equation}
Its solution again is a Gaussian with mean~(\ref{eq:mean:ss}) and variance
$\sigma^2=(2/\beta)\mean{Q\hk}$.

{\sl Relation to other fluctuation theorems. --} Finally, we discuss the
connection between the new integral fluctuation theorem~(\ref{eq:ft}) and the
Hatano-Sasa relation~(\ref{eq:hs}) as well as the Jarzynski-like
relation~(\ref{eq:jar}). For a transition along $\ctrl(\tau)$ the latter two
involve the dimensionless functional
\begin{equation}
  Y[x(\tau);\ctrl(\tau)] \equiv \IInt{\tau}{0}{t}
  \dot\ctrl(\tau)\pd{\phi}{\ctrl}(x(\tau),\ctrl(\tau)),
  \label{eq:Y}
\end{equation}
for which Hatano and Sasa show $\mean{\exp[-Y]}=1$~\cite{hata01}. With the
boundary term
\begin{equation}
  \Delta\phi\equiv\phi(x(t),\ctrl_t)-\phi(x(0),\ctrl_0)
\end{equation}
and $Y=\beta Q\ex+\Delta\phi$ this leads to relation~(\ref{eq:hs}).

Relation~(\ref{eq:jar}) follows by applying the same reasoning used for
deriving the original Jarzynski relation~\cite{jarz97,jarz97a,croo00} to the
present case of broken detailed balance. The ratio of the probability of the
trajectory $x(\tau)$ to the probability of the time-reversed trajectory
$\tilde x(\tau)\equiv x(t-\tau)$ under the time-reversed protocol
$\tilde\ctrl(\tau)\equiv\ctrl(t-\tau)$ is given by
\begin{equation}
  R[x(\tau);\ctrl(\tau)] \equiv
  \ln\frac{\pr[x(\tau);\ctrl(\tau)]\,\ps(x(0),\ctrl_0)}
  {\pr[\tilde x(\tau);\tilde\ctrl(\tau)]\,\ps(\tilde x(0),\tilde\ctrl_0)}
  = \beta Q\tot + \Delta\phi.
  \label{eq:R}
\end{equation}
Here, $\pr[x(\tau);\ctrl(\tau)]$ is the probability of a single trajectory
$x(\tau)$ under the protocol $\ctrl(\tau)$ starting in microstate $x(0)$. It
is easy to show $\mean{\exp[-R]}=1$ and hence relation~(\ref{eq:jar}) follows.

\begin{table}[t]
  \centering
  \includegraphics[width=9cm]{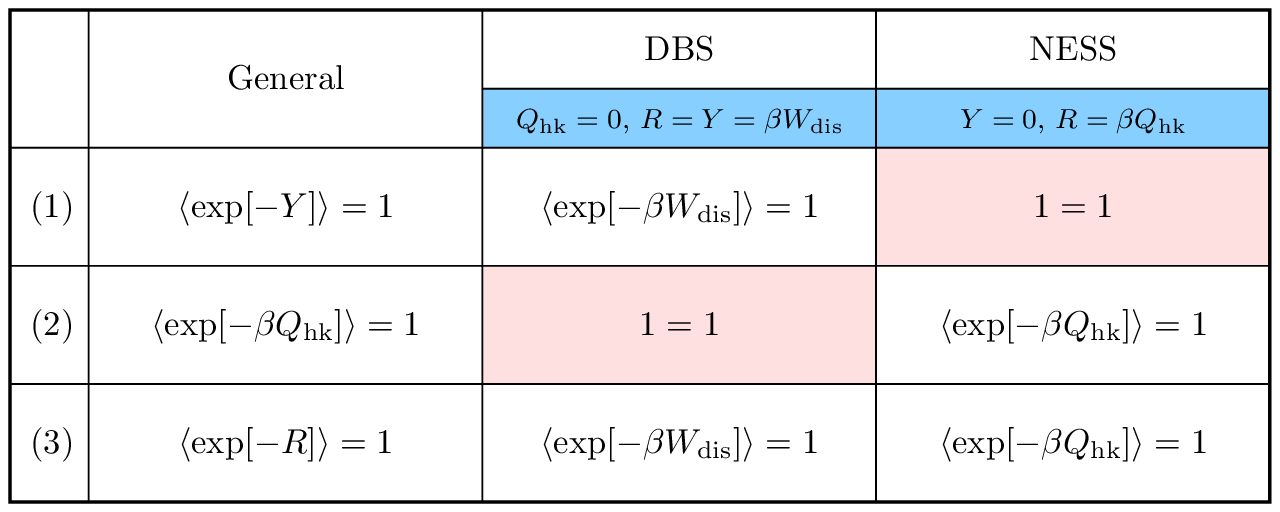}
  \caption{A classification of isothermal nonequilibrium processes
    and the corresponding integral fluctuation
    theorems~(\ref{eq:hs}-\ref{eq:jar}). Here $Y=\beta Q\ex+\Delta\phi$ and
    $R=\beta Q\tot+\Delta\phi$. The second column holds for transitions
    $\ctrl(\tau)$ where for fixed $\ctrl$ the system is in a detailed balanced
    state (DBS). The third column holds in nonequilibrium steady states (NESS)
    with broken detailed balance.}
  \label{fig:overview}
\end{table}

Isothermal nonequilibrium processes are therefore characterized by the two
trajectory dependent functionals $Y$ and $Q\hk$, and a boundary term
$\Delta\phi$. Both functionals and their sum
\begin{equation}
  R = Y + \beta Q\hk = \beta Q\tot + \Delta\phi
\end{equation}
obey an integral fluctuation theorem as summarized in
table~\ref{fig:overview}.  However, there are two special cases in which two
of the three become identical and the third one trivial. First, if for any
fixed $\ctrl$ reached along the transition $\ctrl(\tau)$ the system obeys
detailed balance we have $Q\hk=0$. The stationary distribution then is simply
the Boltzmann distribution with $\phi(x,\ctrl) = \beta[H(x,\ctrl) -
F(\ctrl)]$, where $H(x,\ctrl)$ is the Hamiltonian of the system and $F(\ctrl)$
is the free energy. Inserting this $\phi$ into equation~(\ref{eq:Y}), we find
that $Y$ is the dissipated work
\begin{equation}
  R = Y = \beta[W-\Delta F] \equiv \beta W\dis
\end{equation}
and the relations~(\ref{eq:hs}) and (\ref{eq:jar}) become the Jarzynski
relation~\cite{jarz97}. Second, in a nonequilibrium steady state with
$\dot\ctrl=0$ we have $Y=0$ and $R=\beta Q\hk$, and the
relations~(\ref{eq:ft}) and (\ref{eq:jar}) become identical. In such a steady
state, $R=\beta Q\hk$ can also be interpreted as total entropy production,
which then obeys a detailed fluctuation theorem $P(-R)/P(R)=e^{-R}$ even for
finite times~\cite{seif05a}.

{\sl Summary. --} We have derived and discussed a new integral fluctuation
theorem~(\ref{eq:ft}) for the housekeeping heat valid under arbitrary
time-dependent driving. Even though we focused on one-dimensional motion it is
obvious how to generalize our approach to coupled Langevin equations
describing interacting systems. It would be interesting to test experimentally
all three integral fluctuation theorems and to measure the corresponding
probability distributions, in particular, in nonharmonic time-dependent
potentials, where {\it a priori} one does not expect Gaussian distributions.

\end{document}